    \documentclass[11pt]{article}
    \usepackage[margin=1in]{geometry}
    \usepackage{lmodern}
    \usepackage[T1]{fontenc}
    \usepackage[utf8]{inputenc}
    \usepackage{microtype}
    \usepackage{graphicx}
    \usepackage{booktabs}
    \usepackage{amsmath, amssymb}
    \usepackage{xcolor}
    \usepackage{hyperref}
    \usepackage{multirow} 
    \usepackage{tabularx}
    \usepackage{makecell}
    \usepackage{adjustbox}

    \usepackage{xcolor}
    \usepackage[colorinlistoftodos]{todonotes} 
    
    \newif\ifshowcomments
    \showcommentstrue
    
    \newif\ifbpinline
    \bpinlinetrue

    \hypersetup{colorlinks=true,linkcolor=blue,citecolor=blue,urlcolor=blue}

    \title{Perfecting Human--AI Interaction at Clinical Scale\\ Turning Production Signals into Safer, More Human Conversations}
\author{
Subhabrata Mukherjee\thanks{Correspondence: \texttt{subho@hippocraticai.com}} \ \ \
Markel Sanz Ausin \ \ \
Kriti Aggarwal \ \ \
Debajyoti Datta \and
Shanil Puri \ \ \
Woojeong Jin \ \ \
Tanmay Laud \ \ \
Neha Manjunath \ \ \
Jiayuan Ding \and
Bibek Paudel \ \ \
Jan Schellenberger \ \ \
Zepeng Frazier Huo \ \ \
Walter Shen\and
Nima Shirazian\ \ \ 
Nate Potter \ \ \ 
Sathvik Perkari \ \ \
Darya Filippova \ \ \
Anton Morozov \and
Austin Mease \and
Vivek Muppalla \and
Ghada Shakir \ \ \
Alex Miller \ \ \
Juliana Ghukasyan \and
Mariska Raglow-Defranco \and
Maggie Taylor \and
Herprit Mahal \and Jonathan Agnew \and \\[1em]
\makebox[\textwidth][c]{Hippocratic AI}
}

    \date{\today}

\begin{document}
    \maketitle
    \begin{abstract}
    Healthcare conversational AI agents shouldn't be optimized only for clean benchmark accuracy in production-first regime; they must be optimized for the lived reality of patient conversations, where audio is imperfect, intent is indirect, language shifts mid-call, and compliance hinges on how guidance is delivered. We present a production-validated framework grounded in real-time signals from $115M+$ live patient–AI interactions and clinician-led testing ($7K+$ licensed clinicians; $500K+$ test calls). These in-the-wild cues -- paralinguistics, turn-taking dynamics, clarification triggers, escalation markers, multilingual continuity, and workflow confirmations -- reveal failure modes that curated data misses and provide actionable training and evaluation signals for safety and reliability. 

We further show why healthcare-grade safety cannot rely on a single LLM: long-horizon dialogue and limited attention demand redundancy via governed orchestration, independent checks, and verification. Many apparent “reasoning” errors originate upstream, motivating vertical integration across contextual ASR, clarification/repair, ambient speech handling, and latency-aware model/hardware choices. Treating interaction intelligence (tone, pacing, empathy, clarification, turn-taking) as first-class safety variables, we drive measurable gains in safety, documentation, task completion, and equity in building the safest generative AI solution for autonomous patient-facing care. Deployed across more than $10$ million real patient calls, Polaris attains a clinical safety score of $99.9\%$, while significantly improving patient experience with average patient rating of $8.95$ and reducing ASR errors by $50\%$ over enterprise ASR. These results establish real-world interaction intelligence as a critical -- and previously underexplored -- determinant of safety and reliability in patient-facing clinical AI systems.
    \end{abstract}

\clearpage
\tableofcontents
\clearpage

    \section{Introduction: Real-World Challenges Beyond Static Benchmarks}
\label{sec:introduction}

    Static medical QA benchmarks such as MedQA (USMLE), MedMCQA, PubMedQA, MultiMedQA, and MMLU-clinical subsets ~\cite{medqa,medmcqa,pubmedqa,multimedqa,mmlu} have pushed the field forward by making evaluation scalable and comparable. But this progress has also created a blind spot: the community increasingly optimizes for clean tasks on clean data, then assumes those gains will translate to safe patient-facing conversations. In practice, a live patient call is not a benchmark question. It is speech, not text; it is noisy, not curated; it is emotionally and socially situated; and it is tightly coupled to downstream actions -- scheduling, monitoring, documentation, benefits, escalation, and follow-up. The gap is not subtle. If we want autonomous patient–AI interactions to be safe and reliable, we must learn from the conditions we actually deploy in, not extrapolate from offline leaderboards.

At clinical scale, production conversations exhibit failure modes and opportunity signals that rarely appear in curated datasets. They include acoustic and paralinguistic cues (hesitation, breath, distress markers), turn-taking dynamics and micro-timing, ambiguity and indirect answers, and multilingual continuity with mid-call switching. They also include both system-level and multi-turn feedback that static datasets do not provide: whether an appointment was actually booked given API confirmation, whether a benefits quote can be verified from the source, whether an HRA (Health Risk Assessment) form requires further clarification from the user for reconciliation, and whether escalation was appropriate given the interaction trajectory. These signals are messy—but they’re also information-rich. A production-first approach leverages these signals to capture governed telemetry, surface where systems fail, and convert those patterns into concrete architectural and alignment solutions.

This paper argues for four design principles that follow directly from what live patient calls demand.

{\bf First, real-world signals matter as much as clean accuracy.} A benchmark mindset rewards single-shot correctness on pristine inputs. Live calls require robustness to speech variability and real-time interaction, and they reward repair: knowing when to ask a targeted clarification, how to confirm critical entities, and how to keep the conversation moving without overconfident assumptions. The goal shifts from “correct answer” to “safe completion”: the patient understood, the workflow succeeded, and uncertainty was handled transparently.

{\bf Second, healthcare-grade safety cannot come from a single LLM.} Traditional safety engineering relies on redundancy because complex systems fail in multiple ways. LLMs add a special twist: long-horizon dialogue strains attention and increases the chance of drift, omissions, and misplaced confidence as context grows. A single monolithic model -- no matter how capable -- becomes a single point of failure. We therefore treat safety as a system property, achieved through independent checks, verification, and governed orchestration across components that can catch each other’s misses.

{\bf Third, many “reasoning errors” are really input errors.} In voice-first care, upstream uncertainty is often the root cause: a misheard medication, a swapped digit in a vital sign, background speech mistaken for the patient, or a clipped utterance that changes meaning. If we only improve downstream reasoning, we may simply become more fluent at rationalizing incorrect inputs. Achieving reliable behavior requires vertical integration into the speech stack -- contextual ASR, robust short-utterance handling, clarification/repair mechanisms, and ambient speech control -- so the model reasons over faithful representations of what the patient actually said.

{\bf Fourth, how you deliver care changes outcomes.} In domains like coding, math, or paper writing, tone and pacing rarely affect task correctness. In healthcare, delivery affects disclosure, trust, and compliance. The same instruction can either motivate follow-through or trigger disengagement depending on empathy, turn-taking, and pacing. We therefore elevate interaction micro-skills -- tone calibration, trajectory control, clarification behavior, and conversational timing -- from “nice-to-have UX” to first-class safety variables.

\begin{table*}[htbp]
\centering
\setlength{\tabcolsep}{6pt}
\renewcommand{\arraystretch}{1.25}
\caption{Production-grade clinical intelligence from real-world conversational signals.
Evaluations on live patient interactions and simulated conversations with clinicians show that Polaris 4 reduces clinically relevant errors while improving interaction quality, longitudinal continuity, and responsiveness. The results illustrate the paper's central claim: production-grade clinical intelligence is achieved by learning from real-world interaction signals and embedding them into system-level design, not by optimizing isolated model accuracy alone. }
\begin{adjustbox}{max width=\textwidth}
\begin{tabular}{llcccc}
\toprule
\multicolumn{2}{c}{\textbf{Evaluation Dimensions}} & \textbf{GPT4o} &
\makecell{\textbf{Hippocratic AI}\\\textbf{(Main Model only)}} &
\makecell{\textbf{Hippocratic AI}\\\textbf{(Polaris 4 Constellation)}} \\
\midrule

\multicolumn{5}{c}{\textbf{Evaluating Error Rate ($\downarrow$) on Human--AI Real Conversations}} \\
\midrule

\multirow{3}{*}{Clinical}
& Labs \& Vitals & 18.0\% & 1.5\% & 0.01\% \\
& Medications & 10.9\% & 3.1\% & 0.01\% \\
& Human Escalation & 15.0\% & 7.4\% & 0.07\% \\

\midrule
\multirow{2}{*}{Speech}
& \makecell[l]{Medical Recognition\\(Clinical ASR)} & 12.8\% & 7.3\% & 7.3\% \\
& \makecell[l]{Clarification \&\\Recovery} & 24.6\% & 7.8\% & 2.0\% \\
\midrule
Scheduling
& Appointment Booking & 23.1\% & 13.7\% & 0.1\% \\

\midrule

{Documentation}
& Form Fill & 64.6\% & 15.0\% & 0.5\% \\\midrule
Interactive Voice Response & IVR Navigation & 49.7\% & 18.0\% & 18.0\% \\

\midrule\midrule
\multicolumn{5}{c}{\textbf{Evaluating Win Rate ($\uparrow$) on Simulated Conversations}} \\
\midrule

\multirow{5}{*}{\makecell[l]{HEART\\(Emotional Support\\Dialogue Benchmark)}}
& Conversation Naturalness & 40.9\% & 79.1\% & \multirow{6}{*}{Main model only evaluation} \\
& Empathetic Intelligence & 41.1\% & 78.7\% & \\
& Emotional De-escalation & 50.7\% & 77.2\% & \\
& Likeability \& Engagement & 46.0\% & 85.9\% &  \\
& \makecell[l]{Instruction \& Task-\\following} & 63.9\% & 71.3\% & \\

Multi-call Memory
& Longitudinal Contextualization & 52.0\% & 92.0\% & \\

\midrule
\midrule
Main Model Latency
& Time-to-first-token (TTFT) & 500ms & 400ms & -- \\

\bottomrule
\end{tabular}
\end{adjustbox}

\label{tab:human_ai_evaluation}
\end{table*}

{\bf Key result.} Table~\ref{tab:human_ai_evaluation} aggregates results from multiple evaluation regimes, each matched to the subsystem being measured: retrospective audits of live patient calls for clinical error rates, clinician-validated simulations for interaction quality, and on-policy production measurements for latency and workflow execution. Not all dimensions apply to all model configurations; in particular, Polaris 4 Constellation metrics reflect system-level orchestration beyond a single conversational model. Clinical evaluation protocols are detailed in Section~\ref{sec:evaluation} with description of the sub-tasks for labs, vitals and medications outlined in the first Polaris technical report~\cite{mukherjee2024polaris}. Speech and orchestration evaluation outlined in Section~\ref{sec:speech} and Section~\ref{sec:orchestration}, respectively. Conversational intelligence and empathy evaluation is based on the HEART benchmark~\cite{Iyer2025HEART} discussed in Section~\ref{subsec:heart}. Multi-call memory for longitudinal personalized interactions and latency optimization discussed in Section~\ref{subsec:mcm} and Section~\ref{sec:performance}, respectively. We consider GPT4o as the baseline. GPT-4o (and the GPT-4o realtime variants / family) has been one of the most widely adopted default choices for voice agents, especially across popular voice-agent platforms (e.g., \href{https://vapi.ai}{VAPI}) and Realtime API integrations.

{\bf Outline.} The remainder of the paper shows how these principles translate into a deployable framework. Section~\ref{sec:system} introduces the Polaris safety constellation, where a core conversation model is assisted by specialist models and verifiers, coordinated through governed orchestration rather than a single-LLM decision path. Section~\ref{sec:intelligence} details interaction intelligence: trajectory-aware tone and pacing, empathy-driven dialog control, turn-taking and latency budgeting, and continuity across calls -- framed specifically as safety-relevant behavior in patient communication. Section~\ref{sec:speech} moves upstream to speech understanding for the real world, including contextual ASR and targeted clarification that reduce clinically meaningful input errors before they become downstream failures. Section~\ref{sec:performance} describes the performance and serving constraints that make real-time voice AI possible (and why latency is itself a safety constraint). Section~\ref{sec:orchestration} covers workflow-grounded verification for scheduling, policy quoting/RAG, and documentation reconciliation, emphasizing post-condition checks against sources of truth. Section~\ref{sec:multilingual} addresses multilingual continuity and equity, including mid-call switching and dialectal variability. Sections~\ref{sec:safety},~\ref{sec:evaluation} and \ref{sec:impact} then describe how these components are governed for clinical safety, evaluated at scale using an RWE-LLM approach, and validated through operational and clinical impact in deployment. We discuss related work in Section~\ref{related_work} on traditional static and offline evaluation for both clinical and non-clinical component of AI frameworks, speech understanding and voice systems as well as inference stack and architectural infrastructure.

    \section{System Overview: The Polaris Safety Constellation}
\label{sec:system}

Polaris employs a constellation of specialized LLMs and signal-processing engines surrounding a core conversation model (see our Polaris constellation architecture\footnote{\href{https://hippocraticai.com/research/}{https://hippocraticai.com/research/}}).

\subsection{Core and Specialists}
The constellation comprises of:
\begin{itemize}
    \item A core model that handles dialogue and policy-constrained reasoning.
    \item Over {\em thirty} supervisor models specialized for providing context and reasoning across tasks like medication identification and stoppage, overdose, condition-specific disallowed OTC's, identity verification and compliance, labs and vitals, escalation decisioners that gate high-risk cases, etc. 
    \item Online and offline verifiers that check retrieval and reasoning chains for tasks like structure documentation (HRAs, follow-ups, policy and benefits). Specialists run in two regimes: synchronous steerable guidance and asynchronous ``deep thinking" interleaving that pause-and-verify before sensitive actions.
\end{itemize}

\subsection{Governed Orchestration}
A tool-call layer executes actions (e.g., schedule appointments, transfer calls, send SMS) with governance: preconditions, input validation, and post-conditions (state checks). For instance, a scheduling online checker can query the scheduler to confirm bookings and repair mismatches in-call, while an offline reconciliation model aligns documents with full-call conversational context as opposed to the online one that has access to only partial transcript.
    \section{Interaction Intelligence}
\label{sec:intelligence}

\subsection{Tone Adaptation and Trajectory Control}
Polaris learns trajectory-aware dialog control: adjusting depth and pace to the patient’s signals (reassurance vs. urgency) and employing assertiveness appropriately. EQ features -- such as reading between the lines, picking up on unspoken concerns, and supporting patients who struggle to finish a thought -- improve rapport and make the interaction feel smoother and more attuned~\cite{human_touch,empathetic_intelligence,empathy_in_action}.

In addition to pacing and depth, Polaris continuously adjusts its tone to reflect where the patient is emotionally within the interaction. It softens its language when a patient sounds overwhelmed, becomes more direct when clarity is needed, and maintains steady warmth during sensitive disclosures. These shifts are subtle and unfold over the course of the dialog, helping the patient feel understood without drawing attention to the adaptation itself. By aligning tone with trajectory in this way, Polaris supports smoother conversations, reduces friction during stressful moments, and strengthens the feeling of being guided rather than instructed.

\subsection{Dynamic Conversations, Powered by Empathy}

Polaris tunes for trajectory-aware dialog, adapting depth, tone, and assertiveness to patient needs. Emotional skills -- empathy, reading between the lines, infinite patience, and non-judgmental rapport—build trust, while motivational interviewing promotes adherence. Voice enhancements from professional actors deliver warmth and clarity. These align with empathetic-intelligence principles, increasing comfort in confiding and extended engagement during calls.

Beyond these emotional capabilities, Polaris dynamically adjusts its conversational style to match the patient’s evolving affect and communication patterns. The model blends contextual cues -- urgency, hesitations, distress markers, verbosity, background noises, and lexical uncertainty -- to determine whether to lean into a faster, more directive mode or a slower, warmer, more reflective stance. Style adaptations include adopting a concise clinical tone for medication, dose clarification, or insurance details; shifting into a gentler cadence during emotional overwhelm; or maintaining urgency when the patient signals time pressure or confusion. It also modulates turn length and reasoning depth to respect cognitive load, speech difficulty, or fatigue.

When suitable, Polaris introduces light humor -- never distracting, always calibrated -- to ease tension, restore comfort, or simply keep the interaction human and warm. For patients who express themselves through longer narratives, Polaris maintains a calm, unhurried presence: listening fully, allowing space, mirroring emotions, and responding with steady patience while gently steering the dialogue toward what will help them most. These shifts occur fluidly across turns, preventing the drift or personality collapse seen in single-prompt systems and ensuring the agent remains coherent, stable, and aligned with the patient’s communicative needs.

Patient preferences expressed during the conversation -- such as a desire for brevity, detailed explanations, more encouragement, or a casual tone -- are integrated dynamically into the dialog trajectory. The system continually revises its stance based on new signals, supporting real-time attunement without compromising safety or clinical grounding. Polaris also balances task progression with emotional pacing, avoiding premature reassurance and ensuring the patient feels heard before transitioning to next steps or instructions.

Together, these adaptive behavioral adjustments make interactions feel natural, personalized, and emotionally calibrated. By continuously matching its style to the patient’s psychological and practical needs, Polaris meets patients where they are, sustains rapport across multi-turn settings, and strengthens trust throughout the entire call.
\subsubsection{Benchmarking Interaction Intelligence}
\label{subsec:heart}
\begin{figure}[htbp]
    \centering
    \caption{\textbf{Latency–HEART Elo landscape across models.}
HEART is a benchmark~\cite{Iyer2025HEART} that evaluates supportive-dialogue quality across five dimensions (human-alignment, empathic responsiveness, attunement, resonance, and task-following). As the plot shows, most models achieving high HEART Elo cluster in the multi-second time-to-first-token (TTFT) region, where response delays are too slow for natural turn-taking. \textbf{Polaris~4 is a clear outlier}: it matches the supportive-dialogue quality of much larger
frontier systems while maintaining less than 500ms TTFT (ideal for real-time voice conversations), occupying a sparsely populated region 
of the latency–Elo space where both high empathy and real-time responsiveness are simultaneously achievable.\vspace{2em}
}\includegraphics[width=0.92\linewidth]{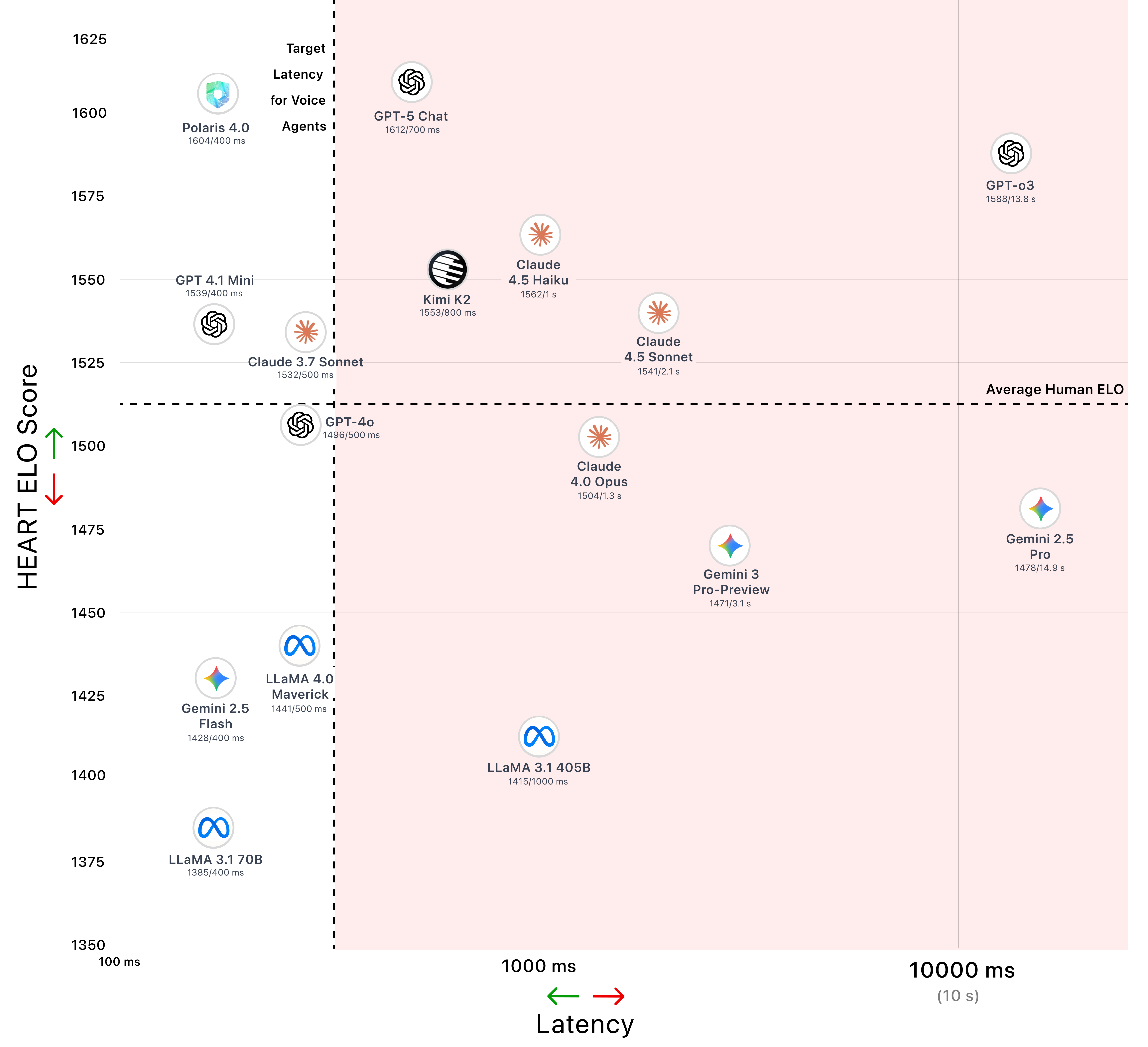}
    
    \label{fig:latency_performance}
\end{figure}
To evaluate Polaris’s interaction intelligence, we benchmark it on \textbf{HEART}~\cite{Iyer2025HEART} -- a recent framework designed specifically to measure supportive, emotionally attuned behavior in multi-turn dialogue. Unlike factual QA or reasoning benchmarks, HEART focuses on the interpersonal dimension of conversation: whether a model responds like a thoughtful, attentive human supporter who listens, calibrates tone, and helps the seeker move forward. HEART evaluates responses along five dimensions grounded in communication science: \textbf{Human Alignment} (natural tone and phrasing), \textbf{Empathic Responsiveness} (affective acknowledgement), \textbf{Attunement} (tracking the seeker’s specific details and emotional signals), \textbf{Resonance} (forward momentum and relevance), and \textbf{Task-following} (respect for safety and role boundaries). These axes jointly capture the micro-skills that shape high-quality emotional support and offer a structured way to measure the kinds of conversational behaviors Polaris is designed to portray.

HEART provides a natural testing ground for Polaris because many of its architectural and alignment choices explicitly target the abilities HEART measures. Polaris’s trajectory-aware control allows the model to shift pacing, framing, and emotional depth across turns, mirroring how human supporters adjust as the conversation unfolds. Its tone-adaptive mechanisms 
-- softening during overwhelm, becoming more direct when clarity is needed, and maintaining warmth during sensitive disclosures -- support both \textbf{Empathic Responsiveness} and \textbf{Human Alignment}. Similarly, Polaris’s clarifying-question heuristics and reflective summarization behaviors strengthen \textbf{Attunement} by grounding the response in the seeker’s specific concerns rather than generic reassurance. HEART’s multi-dimensional scoring captures these competencies in a way that single-turn or sentiment-focused benchmarks cannot, making it well-suited for evaluating interaction-sensitive systems like Polaris.

On HEART, Polaris outperforms other models under ideal latency targets for real-time voice conversations as shown in Figure~\ref{fig:latency_performance}. Polaris also outperforms models with substantially higher latency and larger effective capacities with test-time compute. 
Polaris’s strongest axes are \textbf{Attunement} and \textbf{Empathic Responsiveness}, reflecting its design emphasis on reading-between-the-lines, emotional calibration, and trajectory-aware adaptation. These scores highlight that Polaris’s alignment toward conversational micro-skills produces tangible improvements in how human judges experience its supportive responses.

A distinctive aspect of Polaris’s evaluation is its latency profile. HEART is text-based, but supportive dialogue is highly sensitive to timing, especially in voice-first contexts. As shown in Figure~\ref{fig:latency_performance}, frontier models that achieve the top HEART Elo scores --  GPT-o3, Gemini~2.5~Pro, Claude~4.5~Sonnet -- typically operate at multi-second time-to-first-answer-token (TTFT) values between 2\,s and 22\,s. Polaris~4 occupies a different part of the quality–latency space. It delivers near-frontier supportive-dialogue performance while maintaining a median TTFT of approximately \textbf{400\,ms}, more than an order of magnitude faster than the slowest frontier models. This speed enables naturalistic turn-taking in synchronous voice interactions, preserving the micro-timing cues essential for perceived empathy, conversational flow, and user comfort. Polaris is one of the only models in the high–Elo region operating at less than 500ms TTFT (ideal for real-time voice conversations), alongside Claude~3.7~Sonnet, and significantly faster than larger frontier models such as GPT-o3 and Gemini~2.5~Pro.

Together, these results show that Polaris not only performs strongly on HEART but does so while meeting the responsiveness requirements of real-time interaction. The benchmark highlights several of Polaris’s strengths -- consistent emotional validation, accurate tracking of conversational details, calibrated next-step guidance -- and reveals how domain-specific alignment can achieve human-preferred supportive behavior. 
HEART thus provides evidence that the interaction-intelligence capabilities engineered into Polaris~4 translate into measurable gains on a rigorous, human-centered evaluation of supportive dialogue.
\cite{polaris4doc,skills,empathetic_intelligence,empathy_in_action,human_touch}

\subsection{Turn-Taking and Latency Budgeting}

In real-time voice agents, turn-taking quality is strongly influenced by how quickly the system responds after the user completes an utterance. Human conversation typically features very short gaps between turns -- often on the order of a couple hundred milliseconds -- across languages, and timing is considered a core constraint on language processing in dialogue. When systems routinely exceed that rhythm, users perceive them as sluggish, interrupt them more, or disengage~\cite{Stivers2009TurnTaking,LevinsonTorreira2015Timing,HeldnerEdlund2010Pauses}.

Voice conversations present a unique challenge from lags resulting from transmission, endpointing (EP) and voice activity detection (VAD) particularly difficult in presence of background noise and speech, automatic speech recognition (ASR) and transcription, intermediate language model (LLM) processing and finally the speech generation (TTS). This is why the usual LLM ``time-to-first-chunk" is not as useful as the ``time-to-first-audio" (TTFA) that matters the most. For voice, perceived responsiveness is best approximated by:

\begin{multline*}
  \text{TTFA} = \text{endpointing/VAD} + \text{ASR finalization} + \text{LLM time-to-first-chunk (TTFT)} +\\ \text{TTS time-to-first-audio} + \text{playout/jitter}      
\end{multline*}

Users don't care when the first token appears; they care when the agent starts speaking -- and whether the gap feels like a normal conversational pause~\cite{Stivers2009TurnTaking,LevinsonTorreira2015Timing}. Overall, it's a budgeting problem. We pick an end-to-end responsiveness target that aligns with conversational expectations, then allocate that budget across endpointing/VAD, ASR, LLM, and TTS, and tool-calls, optimizing the modal tail (P95–P99) via our optimized contextual ASR, KV cache optimization, cache-aware routing, workload specific optimization based on use-case and long-context conversation (Section~\ref{sec:performance}).

\subsubsection{Target Latency for Real-time Voice Conversations}
\label{sec:latency-requirements}
We treat a median LLM time-to-first-token (TTFT) of roughly $500\mathrm{ms}$ as a practical design target for real-time voice interaction. Conversation analysis shows that human turn-taking involves extremely short sub-second gap between one speaker finishing and the next beginning~\cite{SacksSchegloffJefferson1978TurnTaking, Stivers2009TurnTaking, LevinsonTorreira2015Timing}. Human–computer interaction research similarly finds that delays below $1\mathrm{s}$ feel fluid, whereas longer pauses begin to feel disruptive~\cite{Nielsen1993Usability}. Together, these strands motivate a sub-second latency budget if we want voice AI agents to feel conversational rather than transactional.

In deployed systems, however, LLM latency is only one contributor to time-to-first-audio (TTFA). Endpointing and ASR typically consume $150$–$300\mathrm{ms}$, and TTS require another $100$–$200\mathrm{ms}$ before producing the first audio frame. To keep overall TTFA below $\sim$$1\mathrm{s}$ under median conditions, the LLM itself must therefore operate within a few hundred milliseconds. A median TTFT of $500\mathrm{ms}$ is thus a reasonable operating point for an LLM for real-time voice AI.

\subsection{Personalized Longitudinal Interactions and Continuity}
\label{subsec:mcm}
Building on the Patient Continuum framework \cite{sanz2025_multicall}, Polaris introduces a multi-call memory architecture that persistently carries non-EHR contextual information across encounters, enabling more personalized longitudinal interactions without compromising patient privacy. These memories are modular and self-contained pieces of patient-specific information. For example, motivational drivers or long-term goals, designed to improve patient engagement and ultimately support better health outcomes. To ensure privacy and regulatory compliance, all memories are extracted and curated using an LLM, and then stored in a HIPAA-compliant database, where they are dynamically retrieved at inference time to inform each new call. When used appropriately, memories significantly improve patient–AI engagement and conversational naturalness.

To maintain safety and trustworthiness, Polaris employs extensive filtering mechanisms to avoid controversial, sensitive, or clinically inappropriate memory content, ensuring that only relevant and clinically constructive information persists across interactions. Furthermore, because effective use of memories depends not only on retrieval, but also on the conversational model’s ability to integrate them appropriately, the conversational model is explicitly aligned to reason about when and how retrieved memories should shape their responses. We created an LLM-as-a-judge-based evaluation benchmark using simulated conversations, to measure the model's ability to appropriately contextualize prior memories in an appropriate and effective way for each patient. Table \ref{tab:mcm-benchmark} shows the results of different models on the multi-call memory contextualization benchmark.

\begin{table}[t]
    \centering
    \caption{Multi call memory contextualization benchmark results.}
    \label{tab:mcm-benchmark}
    \begin{tabular}{l|ccccc}
        \hline
        \textbf{Task} & \textbf{gpt4o} & \textbf{Best Open Model} & \textbf{o1} & \textbf{Polaris 4} \\
        \hline
        Contextualization quality & 52\% & 57\% & 64\% & 92\% \\
        \hline
    \end{tabular}
\end{table}

The contrast below illustrates the impact of alignment when using a memory such as ``the patient’s primary motivation is to live long enough to see his daughter walk down the aisle in a few years.''

\begin{quote}
\textbf{Unaligned LLM}:
Agent: ``Michael, I understand you don’t like using your blood pressure cuff. However, I want to remind you that you told me your goal is surviving to attend your daughter’s wedding in a few years. I think to meet that goal, it’s essential to routinely monitor your blood pressure.''    
\end{quote}

\begin{quote}
\textbf{Patient Continuum–aligned LLM}:
Agent: ``I get it, and I know you’re in a tough spot Michael. On one hand, the blood pressure cuff is annoying, but on the other hand, you want to improve your health – you have so much to look forward to, and you want to be there for your family! How do you think about this tradeoff?"    
\end{quote}

Note the subtlety in conveying the same intent but in different tones that makes the LLM appear pushy in the first instance and motivational in the second. The aligned model uses the memory gently and empathetically, supporting motivation while preserving the patient’s sense of agency. Together, the memory infrastructure and alignment strategy enable Polaris to deliver longitudinally consistent, personable, and clinically grounded conversations while maintaining strict privacy standards.
    \section{Speech Understanding for the Real World}
\label{sec:speech}

In Polaris, we developed a novel contextual ASR architecture to incorporate multi-turn context to handle short utterances, ambiguities, medical context and noise, with modular engines for background speech isolation, slurred speech understanding, and language switching. Trained on curated medical corpora and synthetic noisy data, it achieves $2\times$ lower word error rate (WER) on clinical tasks while improving on baseline WER compared to SOTA open and closed source models. Features like handling background family discussions or forgetful patients (MCI support) ensure robust performance across diverse environments~\cite{polaris4doc,skills,polaris3}.

\subsection{Contextual ASR Architecture}

We implement a multi-turn contextual ASR. This extends standard speech recognition by conditioning the decoding on \emph{multi-turn dialogue context}. Our system is built on a decoder-only \emph{audio large language model} (audio-LLM) that integrates acoustic representations, textual history, and domain-specific corpora into a single generative framework. This unified design allows the model to resolve ellipsis and references across turns, handle rare domain terminology, and follow task-specific prompts, all while maintaining robustness under noise, accent variability, and spontaneous conversational speech.

At a high level, input audio is encoded by the fine-tuned encoder of Whisper-large-v3-turbo~\cite{radford2023robust}, whose multilingual pretraining provides strong general-purpose speech features. These frame-level states are passed through a Conformer-based projector that performs \emph{depthwise-only} temporal merging, compressing adjacent frames per channel while retaining prosodic cues such as pauses and stress patterns. Self-attention layers provide cross-channel mixing, producing a compact token-like sequence aligned to the LLM embedding space. This enables the decoder to process long utterances efficiently without discarding fine-grained phonetic information.

To incorporate conversational context, we encode recent dialog turns and relevant information and prepend them as prefix tokens during decoding. The unified decoder then jointly attends to context and projected audio, allowing it to maintain entity consistency, disambiguate pronouns, and adapt to user-specific phrasing. Domain fidelity is further strengthened by integrating \emph{domain-specific datasets} -- such as medication lists, clinical forms, and policy snippets -- into the training mixture. These corpora expose the model to rare drug names, structured numeric expressions, and other domain-relevant patterns, improving recognition without relying on additional architectural components. Finally, user specific contextual biasing is further taught to the model via synthetic data augmentation using simulated user profiles with personal biasing, such as date of birth, know medication names for patient, addresses, etc. Jointly training with personal biasing context and conversational context allows the model to learn to map the relationship between conversational and personal biasing information: therefore significantly improving performance on real world use-cases of understanding DOB, addresses and proper nouns, often mistranscribed by traditional ASR systems.

\begin{quote}
\textbf{Actual Address}: 1100 Geary Blvd at Geary Blvd \& van Ness Ave intersection.
\end{quote}

\begin{quote}
\textbf{Transcription Without Personalized User Context}:\\[4pt]
\textbf{Agent}: ``Could you please confirm your address?''\\
\textbf{User}: ``Yeah its at the intersection of Gear Boulevard \& Vans Ave at 1100 Gear Boulevard.''
\end{quote}

\begin{quote}
\textbf{Transcription With Personalized User Context}:\\[4pt]
\textbf{UserContext}: \{name: "User Name"; dob: {01/01/1970}; address: "1100 Geary Blvd at Geary Blvd \& van Ness Ave intersection."\} \\
\textbf{Agent}: ``Could you please confirm your address?''\\
\textbf{User}: ``Yeah its at the intersection of Geary Blvd \& van Ness Ave at 1100 Geary Blvd``
\end{quote}

Training follows a two-stage curriculum. First, an alignment stage trains only the projector to match token-affinity distributions between projected audio and ground-truth text, improving the cross-modal mapping prior to end-to-end optimization. Second, the full audio-LLM is fine-tuned with an autoregressive ASR objective, using LoRA on the LLM and updating only the upper layers of the audio encoder for compute efficiency and stability. Noise augmentation (clinic, transit, TV backgrounds) and SpecAugment~\cite{SpecAugment} improve robustness, while a staged curriculum gradually increases contextual complexity and domain exposure, leading to stable learning for multi-turn scenarios.

\subsection{Robust Short Audio Transcription}
We introduce Single Word Correction (SWC) to mitigate a common failure mode in clinical ASR: utterances that produce a one-token transcript. These cases are frequent in patient interactions because many responses are brief affirmations, negations, or short values (e.g., ``okay'', ``yes'', ``no'', ``sure'', numerals), and are disproportionately prone to phonetic confusions such as ``no'' vs. ``now'' or ``five'' vs. ``fine''. When the primary ASR outputs a single word, SWC triggers a secondary verification step that expands the hypothesis set to a small confusion list of phonetically similar candidates. A separate model, conditioned on the broader conversational and call context, then re-scores these candidates to select the most contextually consistent interpretation, acting as an additional guard against mis-transcription while adding only $100$ milliseconds to the ASR latency for single words. Across the evaluated recordings, SWC reduced single-word transcription errors from $2.4\%$ to $0.2\%$ (Table \ref{tab:asr}), yielding a substantial improvement in reliability for short patient replies that often convey critical clinical information.

\subsection{Targeted Clarification and Recovery}
To further improve robustness in real-world calls, we implement Targeted Clarification \cite{skills}, a fallback mechanism for residual ASR errors, incomplete patient responses, or out-of-context inputs caused by background speech and overlapping talk. When the system detects uncertainty — e.g., low ASR confidence, conflicting or implausible entities given domain priors (medications, dosages, names, identifiers), or semantic mismatch with the recent dialogue state — an uncertainty-aware clarifier is triggered. Rather than issuing generic ``please repeat'' prompts, the clarifier generates minimal, high-yield follow-ups that target the most likely confounder (for example, confirming a medication name versus dose, or disambiguating identifiers). This design keeps the conversation natural while ensuring that clinically salient information is confirmed before downstream actions, effectively completing the guardrail stack after SWC.

\subsection{Empathetic Voice}
For Polaris, we redesigned the voice experience using a hybrid human–AI pipeline. We first collected a studio-quality recordings from a professional voice actor, covering clinically relevant interaction types that our agents commonly handle (e.g., pre-procedure reassurance, recovery acknowledgment, and step-by-step explanations of complex medical information). The recordings were structured to elicit consistent prosody, affective range, and context-appropriate speaking styles under controlled conditions.

We then applied voice conversion and normalization methods to reduce variability between sessions and to align timbre, prosodic patterns, and speaking rate with a target synthetic voice profile. The resulting voice model is calibrated to maintain stable acoustic characteristics while retaining human-like expressiveness, yielding a digital voice that prioritizes warmth, clarity, and affect-aware delivery appropriate for healthcare conversations. This process was tested and calibrated for English and later replicated in other languages like Spanish and Arabic. Even for a target language e.g., Arabic the voice was tuned for different dialects like Hejazi, Khaleeji (sub- Emirati), Modern Standard Arabic, etc. 

Audio-based evaluations, specifically involving voice, are quite subtle and nuanced compared to text-based evaluations. We first use a synthetic evaluation using Gemini $2.5$ Pro as the judge, based on prior work~\cite{audiojudge}, using multi-attribute rubrics like empathetic tone, naturalness, warmth, clarity, engagement, overall effectiveness, etc.  Across multiple evaluations, the new voice was selected as better in two-thirds of trials, representing an estimated $30$ percentage-point improvement in preference over the production baseline. Most of the wins were concentrated in utterances that demanded high emotional variability. We additionally conducted a human preference study on a sampled set of audio pairs to validate the LLM-based evaluation. Human raters showed strong agreement, with high inter-rater reliability (IRR), indicating that the preference for the new voice is robust to evaluator choice and not an artifact of the automated rubric scoring.

\subsection{ASR Performance Summary}
Tables~\ref{tab:asr_wer_internal} and \ref{tab:asr_wer_open} show word error rate (WER) on internal evaluation datasets and Open ASR evaluation benchmark. 
Table~\ref{tab:asr} reports component-level error rates from real-world calls. 
Together, they indicate substantial quality gains and markedly better tail latency.

\paragraph{WER.}
On internal evaluations against state-of-the-art enterprise ASR, Polaris lowers WER from $6.47$ to $5.92$ on general domain data (absolute $-0.55$, $\sim\!8.5\%$ relative) and more than halves WER on medical domain data from $15.69$ to $7.76$ (absolute $-7.93$, $\sim\!50.5\%$ relative). The larger domain-specific gain is consistent with our training that incorporates targeted corpora. On the Open ASR benchmark, Polaris is \emph{competitive across diverse conditions}: it leads on SPGISpeech (\textbf{1.76}, best among the listed models), is close on LibriSpeech-Clean ($1.55$; $+0.12$ to the best).
Overall, results suggest that our method’s strengths on domain-critical terminology and clean/read speech carry over to several public domain datasets.

\begin{table}[h]
\centering
\caption{ASR Word Error Rate (WER; lower the better) on internal general and medical domain datasets.}
\begin{tabular}{lcc}
\toprule
\textbf{Model}& \textbf{General} & \textbf{Medical}    \\
\midrule
SOTA Enterprise ASR   & 6.47 & 15.69 \\
Polaris Contextual ASR   & 5.92 & 7.76 \\
\bottomrule
\end{tabular}
\label{tab:asr_wer_internal}
\end{table}

\begin{table}[h]
\centering
\caption{ASR Word Error Rate (WER; lower the better) on the Open ASR benchmark (\url{https://huggingface.co/spaces/hf-audio/open_asr_leaderboard})}.
\begin{tabular}{lccccccc}
\toprule
\small
\textbf{Model}& \textbf{AMI} & \textbf{GS} & \textbf{LS Clean} & \textbf{LS Other} & \textbf{SPGI} & \textbf{Tedlium} & \textbf{Voxpopuli}    \\
\midrule
canary-qwen-2.5b    & 10.19 & 9.43  & 1.61  & 3.1   & 1.9   & 2.71  & 5.66 \\
granite-speech-3.3-8b&8.98  &10.19  & 1.43  & 2.86  & 3.91  & 3.4   & 5.71 \\
Polaris w/ Contextual ASR           & 12.36 & 9.97  & 1.55  & 3.27  & 1.76  & 3.38  & 5.91 \\
\bottomrule
\end{tabular}
\label{tab:asr_wer_open}
\end{table}

\paragraph{Quality improvements by engine.}
Across all engines, Polaris w/ contextual ASR shows large absolute and relative error reductions compared to prior systems. On average (macro over rows), this corresponds to an \emph{85\% relative reduction} in component error. In particular, the large drops for \emph{Entity Transcription} align with our \emph{domain-aware training} -- we incorporate domain corpora (drug dictionaries, forms, policy snippets) during training -- along with curriculum staging and numeric normalization heuristics.

\begin{table}[h]
\centering
\caption{Error rates (lower the better) for real-world calls across different ASR systems used in Polaris.~\cite{polaris3}}
\begin{tabular}{lcc}
\toprule
\textbf{Engine} & \textbf{SOTA Enterprise ASR} & \textbf{Contextual ASR} \\
\midrule
Background Noise Isolation & 9.3\% & 2.3\% \\
Speech Detector (primary speaker) & 15.0\% & 2.4\% \\
Single-Word Recognition & 2.4\% & 0.2\% \\
Entity Transcription (meds/numbers) & 4.2\% & 0.5\% \\
Clarification Engine (misunderstandings) & 16.3\% & 2.0\% \\
\bottomrule
\end{tabular}
\label{tab:asr}
\end{table}

\paragraph{Latency: mean and tails.}
Polaris w/ contextual ASR reduces mean latency 
by $\sim\!15.7\%$ compared to the SOTA enterprise ASR we used in the prior Polaris versions. The largest gains are in the tail by as much as $3.1\times$ latency reduction at P99. 
For completeness, we note that the contextual ASR uses standard decoder \emph{KV caching} and \emph{prefix caching} to avoid recomputation as discussed in the next section.

\section{Performance That Powers Real-Time Care}
\label{sec:performance}

In clinical phone conversations, latency is a safety constraint rather than merely a quality of service metric. Interruptions and ``dead air'' can degrade patient trust and obscure critical diagnostic signals. Polaris 4 achieves a $40\%$ reduction in end-to-end latency at P99 compared to prior versions (Polaris 2.0)~\cite{mukherjee2025polaris2}. This reduction is driven by three primary architectural optimizations:

\begin{itemize}
    \item \textbf{Model Distillation via Layer Pruning:} We derive a 300B parameter generalist backbone from a 405B teacher model using depth-pruning techniques, preserving clinical and reasoning capabilities while significantly increasing token throughput.
    \item \textbf{Memory-Optimized Hardware:} The transition to H200 GPUs provides the High Bandwidth Memory (HBM3e) necessary to support larger batch sizes and persistent KV caches for long-context clinical sessions.
    \item \textbf{Latency-Aware ASR:} A custom ASR engine trained specifically for clinical telephony achieves a 50\% reduction in Word Error Rate (WER) while running $3.1\times$ faster at P99 compared to enterprise ASR.
\end{itemize}

\subsection{Distillation and Model Sizing}

We employ a capability-preserving distillation technique to speedup the main conversation model for real-time voice while maintaining the clinical abilities inspired by recent findings on the inefficiency of deeper layers in Large Language Models \cite{gromov2024unreasonable}.

Gromov et al~.\cite{gromov2024unreasonable} show that in many large transformers, the upper layers contribute disproportionately little to next-token prediction or in-context reasoning, and that the model’s effective computation saturates well before the final blocks. Their work demonstrates that deeper layers often become feature-redundant, exhibit vanishing influence on outputs, and can even degrade performance when retained. Motivated by these findings, we adopt a pruning-then-healing paradigm rather than standard logit-matching distillation.

We identify and remove redundant high-level blocks from the Llama-3.1-405B teacher to construct a 300B student, preserving the “useful compute frontier” while discarding layers that contribute negligible marginal signal. We then apply a continued pre-training (healing) phase to re-align internal manifolds and restore cross-layer coherence, mitigating representational collapse after pruning. As shown in Table~\ref{tab:perf_comparison}, this reduction yields a $30\%$ improvement in request throughput and a significant reduction in Time Per Output Token (TPOT) at P99 (from $266.51$ ms to $117.69$ ms), which is critical for preventing voice latency drift while preserving parity in performance.

\begin{table}[h]
\centering
\caption{Throughput Comparison (Higher is Better): 300B Pruned Model vs. 405B Teacher. The 300B student maintains high throughput with significantly lower tail latency (P99 TPOT). Compared to external provider benchmarks, we have the fastest throughput for serving 405B models.}
\label{tab:perf_comparison}
\begin{tabular}{lcc}
\toprule
\textbf{Metric} & \textbf{300B (Student)} & \textbf{405B (Teacher)} \\
\midrule
Request Throughput (req/s) & 14.31 & 10.96 \\
Input Token Throughput (tok/s) & 2888.49 & 2211.29 \\
Output Token Throughput (tok/s) & 3050.76 & 2335.36 \\
\bottomrule
\end{tabular}
\end{table}

\subsection{Cache-Aware Routing Architecture}

Autoregressive decoding in multi-turn applications is typically bound by memory bandwidth. In standard stateless load balancing (e.g., Round Robin), sequential requests from the same conversation session ($\mathcal{S}$) are distributed stochastically across the inference cluster. This forces the target node to recompute Key-Value (KV) states for the entire history $H_{t-1}$ at every turn $t$, resulting in a prefill latency that scales linearly with conversation depth: $O(|H_{t-1}|)$.

\subsubsection{Deterministic Routing and Cluster Health}

To eliminate redundant computation, we implemented a deterministic routing layer utilizing the Kong API Gateway. We employ consistent hashing on the session identifier (\texttt{call\_id}) to enforce ``sticky routing,'' ensuring that all sequential turns $t$ within a session $\mathcal{S}$ are routed to the same inference node ($\text{node}(t) = \text{node}(0)$). This locality is strictly maintained to maximize cache hit rates, provided the cluster topology remains stable.

To ensure strict adherence to latency SLAs, we augment this routing logic with an active health-check protocol running at 5-second intervals. This high-frequency probing allows the load balancer to identify and preemptively remove degraded or unreachable nodes from the consistent hash ring \textit{before} request dispatch. By preventing requests from queuing on failed nodes, we eliminate head-of-line blocking and timeout-induced latency spikes, ensuring that the P99 inference times 
are maintained even during partial cluster outages.

\subsubsection{KV Cache Persistence and Efficiency}

This routing guarantee allows the inference engine to persist the KV cache in high-bandwidth GPU memory (HBM3e on H200s). For any turn $t > 0$, the system bypasses the prefill phase for $H_{t-1}$, computing attention scores only for the new user utterance and system/tool outputs.

As detailed in Table~\ref{tab:cache_metrics}, this shifts the computational profile from a compute-bound prefill (Turn 0) to a memory-bound decode (Turn $1+$). In the steady state, the Mean Cache Hit Rate (CHR) converges to 96.4\%, resulting in an 18x reduction in estimated prefill latency ($\sim$450ms $\rightarrow$ $\sim$25ms). Critically, this decouples system responsiveness from context length, preventing the ``slowdown'' artifact common in long-context workflows.

\begin{table}[h]
\centering
\caption{Prefix Cache Efficiency Metrics: Cold Start vs. Steady State. By leveraging consistent hashing, the system achieves a 24x context reuse factor, effectively amortizing the cost of the initial prefill across the entire session.}
\label{tab:cache_metrics}
\begin{tabular}{lccc}
\toprule
\textbf{Metric} & \textbf{Cold Start (Turn 0)} & \textbf{Steady State (Avg)} & \textbf{Delta} \\
\midrule
Mean Cache Hit Rate (CHR) & 0.0\% & 96.4\% & +96.4 pts \\
Avg. Re-computed Tokens (Miss) & 2,450 & 128 & -94.8\% \\
Effective Context Reuse & 1.0x & 24.5x & 24x \\
Est. Prefill Latency & $\sim$450 ms & $\sim$25 ms & 18x Faster \\
KV-Cache Memory Eviction Rate & 100\% & $< 1.5\%$ & Stable \\
\bottomrule
\end{tabular}
\end{table}

\subsection{Workload Analysis Across Clinical Domains}

To validate the robustness of our cache-aware routing, we analyzed token distribution across five distinct production workflows, representing a diverse range of clinical complexity. As illustrated in Table~\ref{tab:workflow_performance}, these workloads impose different stress tests on the inference infrastructure:

\begin{itemize}
    \item \textbf{Inbound Scheduling (High-Context RAG):} Workflows such as the \textit{PCP Office Hotline} require retrieving and injecting massive schedule availability blocks and provider directories into the context window. This results in a high ``Cold Start'' volume ($> 8,500$ tokens). However, our routing mechanism ensures that this heavy context is cached, keeping subsequent turn latency low despite the massive prompt size.
    \item \textbf{Discharge Follow-Up (Long-Horizon Dialogue):} The \textit{Inpatient Discharge} workflow represents a ``depth'' challenge, often exceeding $60$ turns as the agent reviews complex post-acute care instructions. The steady-state caching prevents latency degradation even as the conversation history approaches the context window limit.
    \item \textbf{Care Gap \& Welcome Calls (Standard Clinical):} Routine outreach workflows (e.g., \textit{Care Gap Closure}) exhibit a balanced profile, where the system efficiently manages state verification without incurring the re-computation penalties typical of stateless architectures.
\end{itemize}

\begin{table}[h]
\centering
\caption{Profile of Analyzed Production Workloads. We selected five representative workflows to demonstrate cache efficiency across varying context lengths (RAG intensity) and conversation depths (Turn count).}
\label{tab:workflow_performance}
\resizebox{\columnwidth}{!}{%
\begin{tabular}{llcc}
\toprule
\textbf{Clinical Workflow} & \textbf{Workload Characteristic} & 
\textbf{Context Profile} \\
\midrule
\textbf{Inbound PCP Office + Scheduling} & High-Context RAG (Schedule Injection) & 
Heavy Initial Load \\
\textbf{Inpatient + Discharge Follow-Up} & Long-Horizon Diagnostic Dialogue & 
Linear Growth (High Depth, $> 60$ turns) \\
\textbf{Care Gap Closure} & Protocol-Driven Interview \& Longitudinal Follow-ups & 
Balanced (Moderate Depth, $> 50$ turns) \\
\textbf{Insurance Benefits} & Engagement \& Verification & 
Short-Horizon \\
\bottomrule
\end{tabular}%
}
\end{table}

    \section{Improved Orchestration Features}
\label{sec:orchestration}

\subsection{Appointment Scheduling Online Verifier}
Scheduling medical appointments with generative AI is uniquely challenging: it requires extremely high accuracy, minimal hallucination rate, and the ability to handle unpredictable real-world behavior. Patients routinely change preferred times, reject available slots, or seek multiple appointments across different specialties. Many scheduling calls are also clinically rich, involving medications, symptoms, or abnormal labs/vitals, which means the system must distinguish when to book a routine appointment, an acute appointment, and when to advise the patient to seek urgent or emergency care. 2.74\% of our scheduling calls involve patients sharing one or more symptoms they are experiencing. As an example, consider the overview of a call that features a highly complex patient describing multiple recent falls (including hitting her head), severe weakness, osteoporosis, a painful hip she’s afraid is injured, Crohn’s disease with extreme GI symptoms, medication side effects, and clear emotional distress and depression. The Polaris AI agent spends time listening, reflecting her feelings, and untangling a messy history of ER visits, GI care, and failed provider fit, while repeatedly validating how overwhelmed and scared she feels. Instead of treating this as a routine scheduling request, the system recognizes the combination of falls, possible hip injury, head strike, profound weakness, bruising, and mental health strain as high risk, and ultimately routes her to a live team member for more urgent evaluation and care coordination. This preserves clinical urgency and ensures she isn’t left waiting for a standard office visit when her situation needs human judgment and potentially faster intervention.

To ensure reliability in this high-stakes setting, we introduce a hybrid online verifiers' framework composed of rule-based and model-based verifiers. These verifiers monitor the agent’s actions in real time, confirming that proposed appointments actually exist in the scheduling system and immediately correcting errors such as booking inconsistencies or appointment hallucinations. The primary safety target is the hallucinated appointment rate -- the rate at which the system tells a patient that an appointment was booked when it was not.

Across 
thousands of audited scheduling-related interactions, the Polaris AI agent had a hallucination rate of 0.49\%. However, with the online verifier enabled, the scheduling hallucination rate sharply dropped to 
{\bf 0.13\%} allowing the system to self-correct during the course of the conversation. The remaining $0.13\%$ hallucinations were subsequently caught by offline model-based verifiers within minutes, giving operational teams enough time to call the patient back and correct the information. This combination of real-time and near-real-time verification enables medical-grade scheduling performance even in complex, clinically detailed conversations.


\subsection{IVR Navigation, Policy Quoting / RAG}
Agents navigate payer/provider Interactive Voice Response (IVR) systems and quote policies exactly with citations, using a Retrieval-Augmented Generation (RAG) stack built in Polaris 4; policy-quoting accuracy sustains 99.4\% at larger scale~\cite{polaris3}.
The retrieval layer uses an embedding model that is fine-tuned with contrastive learning on a diverse mixture of domains, including payer/provider IVR scripts, policy documents, and durable medical equipment (DME) manuals. During fine-tuning we mine hard negatives across domains, which improves discrimination between closely related policy clauses and enables one unified retriever to serve heterogeneous use cases such as IVR navigation, policy quoting, and DME operating guidance.

Polaris 4 supports a wide variety of document formats commonly encountered in payer/provider operations, including semi-structured tables, FAQ-style Q\&A, long-form manuals, and plan policy PDFs. We design customized indexing pipelines for each document type, including table-aware chunking and header propagation for tabular files, as well as semantic segmentation strategies for long-form manual documents, so that semantically corresponding queries are consistently mapped to the appropriate segment chunks in the embedding space for retrieval. Internal evaluations on policy-quoting workloads show that our customized indexing strategy yields 99.4\% accuracy at scale, which in turn enables agents to quote policies exactly with citations.

To further control hallucinations, Polaris 4 incorporates two LLM-as-judge verifiers around the generator. A \emph{retrieval verifier} first inspects candidate context chunks and filters out passages that are irrelevant to the user’s query, reducing the chance that spurious context will steer the model off-policy. A \emph{generation verifier} then evaluates whether the drafted answer is fully grounded in the remaining retrieved evidence; if unsupported content is detected, the system triggers a constrained revision step. This two-stage verification effectively drives hallucination rates to $0.01\%$ in offline evaluations while preserving the accuracy of policy quote.

\subsection{Documentation Reconciliation / Form Fill}

Conversational LLM systems in task-oriented settings are expected to transform patient–agent dialogues into structured records that can be consumed by downstream workflows such as appointment scheduling and clinical follow-up orchestration. We refer to this end-to-end capability as the \emph{Form Fill} system. In this section, we formalize the Form Fill problem, describe the hybrid online–offline architecture used in our deployment, and summarize its empirical performance under realistic noise and backend constraints.

Form Fill operates over multi-turn, voice-based conversations that are first transcribed by an Automatic Speech Recognition (ASR) system and then processed by an LLM.
Formally, we model each interaction as a dialogue $(D = (u_1, \dots, u_T))$ of transcribed user and agent utterances, and the target output as a structured record \(y = (y_1, \dots, y_K)\) comprising fields needed by downstream systems (e.g., name, phone number, email addresses, contact details, preferences, question responses) together with derived actions such as follow-up questions or appointment requests. 
Each conversation is associated with a \emph{form} template that specifies which pieces of information should be collected. 
We denote the form schema by
\(\mathcal{F} = \{(q_k, \tau_k)\}_{k=1}^K\), where \(q_k\) is the \(k\)-th scripted question and \(\tau_k\) is the type or domain of the corresponding field (e.g., name, phone number, email address, preferences). 
The target output of Form Fill is then a structured record, with each $y_k$ belonging to \(\tau_k\) or a special symbol \(\bot\) indicating that the field is not filled, i.e., $y_k \in \{\tau_k, \bot \}$. 

The core difficulty is that user-provided information is often fragmented, off-script, and revised over time: answers may be implicit, spread across multiple turns, or contradicted and corrected later in the call. 
The questions could be asked in various ways and orders, spread over multiple turns, or completely skipped.
In addition, ASR and LLM interaction introduces its own class of errors, so the LLM must extract correct structured information from noisy text while maintaining alignment between evolving dialog context and the underlying information needs.

\subsubsection{Hybrid Online-Offline Architecture}

To handle these challenges, we implement Form Fill as a hybrid online-offline process rather than a single-pass extractor. 

The online component runs during the conversation and is optimized for responsiveness and script alignment.
During each conversation, it maintains the form questions \(\{q_1, \dots, q_K\}\) from the schema $\mathcal{F}$ defined above and, for each dialog prefix \(D_{1:t}\), performs \emph{question detection} to identify which question, if any, is currently being addressed.
Conditional on a non-null prediction \(\hat{\imath}_t \in \{1,\dots,K\}\), an \emph{answer extraction} component then proposes a candidate value \(\hat{y}^{\text{online}}_{\hat{\imath}_t}\) based on a localized context window around \(u_t\). This coupling between the detected question and extraction context reduces spurious matches and keeps the evolving record aligned with the script as the call unfolds.

After the call completes, an offline phase operates over the full transcript \(D\). A \emph{reconciliation} step re-scans the entire conversation and, for each field \(k\), collects one or more candidate answers together with supporting evidence, producing a second set of candidates \(\{\hat{y}^{\text{offline}}_k\}_{k=1}^K\) that is independent of the online alignment. Because it has access to all turns, this step can recover answers that were missed online (for example, when the user answers a question before it is asked or much later in the call) and flag fields for which multiple, conflicting values were mentioned. 
A subsequent \emph{arbitration} step then combines online and offline candidates, together with other metadata to select a final value \(y_k\) or mark the field as unresolved. 

In effect, the online component provides a low-latency initial record, while the offline reconciliation–arbitration pipeline acts as a consistency check and late-correction mechanism that improves overall accuracy without impacting the live user experience. All reported accuracy numbers in this section are computed over the final post-arbitration record \(y\).

\subsubsection{Edge Cases, Safety Mechanisms, and Downstream Workflows}

In practice, Form Fill must remain reliable in the presence of several recurring edge cases. 
High-stakes fields such as names, phone numbers, email addresses, medication names, and dates are particularly sensitive to ASR errors, tokenization issues, and model hallucinations: a single substitution or transposition can result in an erroneous record. 
To reduce silent failures on these high-stake fields, the system applies a \emph{require confirmation} policy: candidate values are surfaced back to the user in natural language, and only explicitly confirmed values are committed to the structured record. 
This introduces a small interaction cost but substantially lowers the risk of incorrect records.

A second failure mode is \emph{de-synchronization} between the conversation and the underlying form, especially in longer calls with digressions and clarifications. 
If the model over-relies on distant context, it may attach an answer to the wrong question or overwrite a previously correct field with an unrelated value. 
To mitigate this, the online question detection component is constrained to operate over a narrow, recency-weighted context window that is explicitly anchored to the current script state. 
This reduced-context design makes the question detection step more reliable.

Finally, the structured record \(y\) produced by Form Fill drives downstream workflows such as appointment scheduling and follow-up creation. 
In these cases, the system must translate \(y\) into concrete tool calls (for example, constructing API requests to scheduling backends) and ensure that the resulting actions respect constraints such as availability, timing, and basic eligibility rules. 
We therefore treat tool invocation as part of the Form Fill pipeline and apply the same principles of explicit confirmation for high-impact actions, conservative handling of ambiguous inputs, and post-hoc consistency checks before any irreversible operation is executed.

\subsubsection{Evaluation and Empirical Performance}

We evaluate Form Fill at the level of individual fields and end-to-end task outcomes. 
For structured records, we use field-level exact-match accuracy: the fraction of fields whose final value \(y_k\) exactly matches a human-validated label. 
This is directly analogous to slot accuracy in dialogue state tracking. 
For Form-Fill evaluation, we ran human review on a randomly sampled 2.5\% of all production calls, with sampling configured to cover diverse use cases and a mix of long and short conversations and forms with many or few questions. In these audits, human experts re-checked every Form Fill field. 

Across two recent evaluation windows in Q4 2025, the fraction of fields that required correction was $0.117\%$ and $0.17\%$, corresponding to field-level exact-match accuracies of $99.883\%$ and $99.83\%$, respectively. 
Among calls that contained at least one Form Fill error, $60\%$ were attributable to upstream ASR problems, or application orchestration issues, including cases where Form Fill was not configured, and $40\%$ were due to intrinsic detection or extraction errors in the Form Fill model itself. At the call-level, the fraction of audited calls with at least one corrected Form Fill field was $0.58\%$ and $0.44\%$.

Our model's observed accuracy at field-level improved from $98.5\%$ (Polaris 3) to $99.86\%$~\cite{polaris3} in Polaris 4, with a substantial subset of residual errors arising outside the core detection and extraction components.





    \section{Multilingual Continuity and Equity}
\label{sec:multilingual}

Building a safe and equitable phone-based clinical AI assistant requires robustness across linguistic, cultural, and interactional variability. Spoken dialogue introduces cascading sources of error—accents, dialects, code-switching, low-resource language features, and pragmatic differences—that disproportionately affect non-English speakers. These breakdowns can directly translate into safety risks during real-time care. We highlight the key failure modes and our corresponding mitigation strategies below.

\subsection{Accuracy Gaps in ASR/TTS/LLM}

Multilingual speech understanding remains challenging: multilingual ASR systems typically regress in word-error rate (WER) relative to language-specific models. A common failure—such as interpreting the Spanish “sí” (“yes”) as the English letter “C” or the English terms "Sea" or "See"—illustrates how small transcription errors can propagate into incorrect clinical confirmations. Although English ASR achieves low WER, many languages show substantial degradation. Arabic is particularly difficult due to wide phonetic variability, scarce training data, and major dialectal divergence (e.g., Hejazi, Emirati, Khaleeji). Medication terminology further amplifies these issues, with generic Arabic ASR systems often exceeding ~30\% WER on medication names.

To mitigate these errors, our system performs continuous language identification and rapid automatic switching across ASR, TTS, and LLM modules. Rather than assuming a monolingual caller, it maintains a parallel multilingual safety state for English, Spanish, and Arabic, enabling millisecond-level realignment when users code-switch. Medication entities, numerals, and safety-critical context are preserved across language boundaries.

For Arabic, we use multi-ASR ensembling with medication-focused decoding pipelines. Models vote on medication entities using phonetic similarity scoring and transliteration harmonization, substantially reducing the high WER typical of general-purpose systems.

\subsection{Dialectal and Cultural Variability}

Dialectal and cultural differences shape how callers express needs, describe symptoms, and interpret tone. Spanish speakers vary in politeness markers and directive strength across Mexico, Puerto Rico, and U.S. immigrant communities. Arabic dialects diverge to the point that phrases acceptable in Modern Standard Arabic may sound overly formal or confusing in Gulf dialects. These stylistic shifts require dynamic detection and adaptation mid-call to maintain clarity and trust.

\subsection{Mid-Call Language Switching and Multilingual Interference}

Code-switching—especially between English and Spanish or Arabic—is common when callers reference numbers, medications, or insurance/legal terms. Traditional ASR systems often treat this as noise, causing incorrect entity extraction and mismatched safety prompts. Multilingual ASR models can also blend languages, silently injecting cross-lingual synonyms that lead to unsafe LLM inferences. Spanish and Arabic illustrate two critical but distinct cases: Spanish as a high-volume U.S. language with strong bilingual patterns, and Arabic as a lower-resource, highly dialectal language. Our multilingual and dialect-aware safety strategies produced measurable improvements in deployment. We highlight the case of a patient contacted during one of our summary heat-wave safety outreach programs. Although the AI agent began the call by introducing itself in English, the patient responded solely in Spanish, stating that they did not speak any English. Our system recognized the mid-call language shift and automatically switched the ASR, TTS, and LLM components to their Spanish counterparts, and completed the rest of the call in Spanish.

    \section{Uncompromising Clinical Safety}
\label{sec:safety}

Safety remains the cornerstone of the system, with Polaris 4 achieving a $99.9\%$ no-error rate ($0.1\%$ no-harm errors and $0\%$ minor harm, severe harm or death) across all connected calls as shown in Table \ref{tab:safety-performance}, surpassing prior versions and even average human clinician performance for equivalent tasks\footnote{Assessed by Human US Licensed Physicians and US Licensed Nurses}. The clinical escalation system relies on specialized agents—covering labs and vitals, medications, and escalations—with higher accuracy and more up-to-date knowledge than prior versions, which is especially relevant for understanding new medications. These agents collaborate to ask targeted follow-up questions, reducing aggregate over-escalation rates while preserving safety. In Polaris 4, the labs and vitals specialist had an error rate of 0.005\%, the medication specialist 0.01\%, and the escalation specialist 0.07\%, with all errors across agents classified as “no harm”. These results show a significant improvement when compared to Polaris 3, as shown in Table \ref{tab:escalation-rates}.

In August of 2025, the overall escalation rate out of all connected calls—defined as any call requiring immediate transfer to a human or review within 24 hours—was 0.77\%, down from 3.4\% in June of 2025, while the calls that required an immediate transfer were 0.26\%, down from 1.22\%. At the same time, the proportion of connected calls categorized as ``no-harm'' decreased fivefold, from 0.5\% to 0.1\%. This reduction balances autonomy with human oversight, minimizing unnecessary transfers that could overwhelm human teams \cite{polaris4doc,skills,mukherjee2025polaris3} while developing the safest generative AI system for healthcare.

\begin{table}[t]
    \centering
    \caption{Safety Performance of Polaris Model Family and Human Clinicians.}
    \label{tab:safety-performance}
    \begin{tabular}{l|cccccc}
        \hline
        \textbf{Model} & \textbf{Correct Advice} & \textbf{No Harm} & \textbf{Minor Harm} & \textbf{Severe Harm} & \textbf{Death} \\
        \hline
        Polaris 4.0 & 99.90\% & 0.10\% & 0.00\% & 0.00\% & 0.00\% \\
        Polaris 3.0 & 99.38\% & 0.55\% & 0.07\% & 0.00\% & 0.00\% \\
        Polaris 2.0 & 98.75\% & 1.02\% & 0.13\% & 0.10\% & 0.00\% \\
        Polaris 1.0 & 93.23\% & 4.55\% & 1.83\% & 0.32\% & 0.06\% \\
        \hline
        Human Clinicians & 81.16\% & 14.72\% & 4.12\% & 0.00\% & 0.00\% \\
        \hline
    \end{tabular}
\end{table}

\begin{table}[t]
    \centering
    \caption{Escalation rates and error rates for Polaris 3 and 4.}
    \label{tab:escalation-rates}
    \begin{tabular}{l|cc}
        \hline
        \textbf{Metric} & \textbf{Polaris 4.0} & \textbf{Polaris 3.0} \\
        \hline
        Escalation Rate – Overall   & 0.77\%   & 3.40\% \\
        Escalation Rate – Immediate & 0.26\%   & 1.22\% \\
        Error Rate – Lab/Vital      & 0.005\% & 0.06\%     \\
        Error Rate – Medication     & 0.01\%  & 0.02\%     \\
        Error Rate – Escalation     & 0.07\%  & 0.32\%     \\
        \hline
    \end{tabular}
\end{table}
    \section{Evaluation at Scale: RWE-LLM in Practice}
\label{sec:evaluation}

\subsection{Overview of the Evaluation Framework} Our evaluation of Polaris builds on the Real-World Evidence LLM (RWE-LLM) methodology described in the Hippocratic AI Safety Framework \cite{rwellm}. This methodology integrates clinician simulation, on-policy testing, automated LLM-based rater assessments, and retrospective safety reviews. Together, these components allow for continuous validation of safety, reliability, conversational quality, and equity performance. Unlike traditional model evaluation pipelines that rely solely on offline test sets, the RWE-LLM system incorporates evidence from large-scale, production-proximal interactions, enabling more accurate detection of failure modes and contextual performance variation.
\subsection{Clinician Simulation}Clinician simulation serves as the foundation of the evaluation process. More than 7,000 licensed clinicians have participated in simulation efforts to date, generating over 500,000 structured test calls in which each clinician interacts with the AI as they would with a real patient. These interactions capture detailed feedback on clinical reasoning, medication safety, benefit and identity verification, symptom clarification, and escalation accuracy. Prior research demonstrates that clinician-generated labels reliably capture safety-critical judgments in conversational agents \cite{bickmore,zhang}. The resulting label corpus provides training and calibration signals for verifier models, alignment layers, and task-specific guardrails. This simulation pathway allows early identification of confusion patterns, conversational breakdowns, or clinical reasoning deficits before models are exposed to patient-facing environments.
\subsection{On-Policy Evaluation Under Real Conditions}On-policy evaluation complements clinician simulation by testing the system under realistic conversational conditions. These evaluations expose the AI agent to natural user variability, including accent diversity, background noise, partial disclosures, interruptions, and heterogeneous communication styles. Model variants are compared across metrics such as safety-event frequency, escalation accuracy, procedural correctness, benefits verification accuracy, and patient-reported satisfaction. This methodology is especially important for assessing the governed orchestration layer described in the Safety Framework, which enforces preconditions, input validation, and post-condition checks to ensure safe execution of sensitive tasks \cite{rwellm}. On-policy testing is therefore the primary mechanism for detecting safety breakdowns that may not emerge in scripted test environments.
\subsection{Automated Rater Assessment}Automated raters extend evaluation coverage by applying LLM-based scoring systems trained on clinician-generated labels. These raters assess conversational quality, coherence, tone, empathy alignment, motivational interviewing technique, clarification depth, and policy adherence. Automated evaluation enables continuous model monitoring and rapid iteration cycles by scoring every call rather than a small subset. This is particularly important in memory-enabled contexts, where the agent must make contextually appropriate references to prior interactions. Evidence from the multi-call memory study indicates that each additional memory reference extends call duration by approximately 2.47 minutes without lowering satisfaction \cite{sanz2025_multicall}, underscoring the value of rater systems that evaluate memory relevance, appropriateness, and timing.
\subsection{Retrospective Safety Review}Retrospective reviews serve as the final component of the RWE-LLM framework. These reviews examine transcripts, escalation logs, incident reports, and verified safety events to identify emergent or rare error modes. The insights generated from these reviews feed into updates to safety verifiers, escalation logic, tone-modulation strategies, and conversational-policy constraints. As emphasized in the Safety Framework, ongoing retrospective analysis plays a crucial role in ensuring that model evolution remains aligned with clinical expectations and organizational governance requirements \cite{rwellm}. Together, these review processes integrate past experience into future system behavior, forming a closed-loop safety ecosystem.
\subsection{External Validation Through Case Studies}Real-world deployments further validate the RWE-LLM framework. Case studies from large health-plan implementations demonstrate that the evaluation architecture remains robust when applied to high-volume, complex workflows such as benefit verification, care-gap closure, appointment coordination, longitudinal care management, and multi-call continuity \cite{rwellm}. Across these domains, the RWE-LLM system consistently identified failure modes early and guided model updates that improved reliability and safety. This cross-setting consistency supports the framework’s generalizability across clinical, administrative, and preventive-care contexts.
     \section{Operational and Clinical Impact Across Settings}
 \label{sec:impact}
 
\subsection{Overview}Deployments of the AI care agent across administrative, clinical, and preventive-care workflows have generated consistent improvements in operational capacity, engagement, and clinical process reliability. Health systems report increased workflow throughput, more reliable benefits verification, improved care-gap closure, and greater adherence to chronic-care protocols \cite{homepage}. These gains emerge from the agent's ability to handle repetitive communication tasks with high consistency, freeing clinical and administrative staff to focus on cases requiring human judgment.
\subsection{Safety Validation at Clinical Scale}Prior to real-world deployment, the AI care agent underwent the largest empirical safety validation study conducted for healthcare AI to date. This nationwide evaluation encompassed 306,965 unique clinical interactions assessed by 6,527 licensed US clinicians—including 6,294 registered nurses and 233 physicians—spanning all 50 states and the District of Columbia.\cite{rwellm}

The validation framework employed a novel three-tier methodology emphasizing comprehensive output testing rather than input validation. Tier 1 involved direct AI–clinician interaction testing, where participating clinicians engaged with the system and flagged safety concerns. Tier 2 consisted of systematic review by specially trained internal nursing teams who assessed clinical validity of flagged concerns and determined severity levels. Tier 3 provided independent physician adjudication for complex cases, with emergency medicine and primary care physicians delivering definitive clinical judgment on safety implications.

This approach directly addressed fundamental limitations in current healthcare AI evaluation, which typically relies on benchmark testing of hundreds rather than hundreds of thousands of interactions. By testing actual system outputs across diverse clinical scenarios—including routine care coordination, medication management, chronic disease education, and emergency situations requiring escalation—the framework provided validation coverage orders of magnitude greater than traditional approaches.

The validation demonstrated substantial measurable safety improvements across four developmental iterations of the system. Correct medical advice rates progressed from approximately 80\% in baseline testing to 99.58\% (95\% CI: 99.53\%–99.63\%) in the final version.\cite{rwellm} This 19.58 percentage point improvement represents a clinically meaningful advancement achieved through systematic validation-driven development.

Critically, potentially harmful advice was reduced to near-zero levels. Incorrect advice with potential for minor harm declined from 1.89\% (95\% CI: 1.67\%–2.14\%) in early iterations to 0.08\% (95\% CI: 0.06\%–0.11\%) in the final version—a 95.8\% relative reduction. Incorrect advice with risk of severe harm decreased from 0.33\% (95\% CI: 0.25\%–0.44\%) to 0.00\%, representing complete elimination of severe harm risk. Risk-of-death errors, initially present at 0.05\% (95\% CI: 0.03\%–0.11\%), were eliminated entirely in later versions, maintaining 0.00\% rates across subsequent iterations.\cite{rwellm}

These quantitative outcomes establish new empirical benchmarks for healthcare AI safety, demonstrating that systematic validation can achieve safety standards exceeding typical human clinical communication benchmarks. The progressive improvement across system iterations provides the first large-scale empirical evidence that comprehensive pre-deployment validation can ensure AI safety in healthcare settings, challenging the widespread assumption that safety can be inferred from training data quality alone.
\subsection{Impact on Chronic Disease Monitoring}The remote patient monitoring program in nephrology provides one of the clearest illustrations of system-level clinical impact. Among 5,590 older adults across 18 states, a single AI-delivered welcome call more than doubled maximum call duration (205 to 431 seconds), increased verified call rates from 11.9\% to 30.5\%, and increased call-completion rates from 46.2\% to 62.4\% \cite{agnew2025_welcome}. These effects persisted across age, sex, and region, with regression models explaining less than 2\% of variance, demonstrating broad generalizability even in patients aged 75 years and older. Improved engagement led directly to more reliable monitoring of blood pressure, faster follow-up for abnormal values, and improved adherence to chronic kidney disease management protocols.
\subsection{Longitudinal Interaction Quality Through Multi-Call Memory}The addition of multi-call memory further strengthens the agent’s impact across longitudinal care pathways. Memory-enabled conversations were shown to increase behavioral engagement substantially, with each additional memory reference increasing call duration by an average of 2.47 minutes \cite{sanz2025_multicall}. Although satisfaction levels remained stable, the increased depth and continuity of these conversations allowed patients to engage more fully, supporting richer discussions and more coherent interactions across repeated encounters. This type of longitudinal coherence is especially valuable in chronic disease management, where repeated reinforcement of goals and understanding of patient-specific barriers are essential.
\subsection{Reduced Disparities Through Multilingual Preventive Outreach}Preventive outreach results demonstrate the potential of AI agents to reduce disparities in population health. In a multilingual colorectal cancer screening initiative, Spanish-speaking patients—historically exhibiting lower screening rates—showed significantly higher engagement, including connect rates of 69.6\% versus 53.0\% among English speakers, and more than twice the FIT test opt-in rate (18.2\% vs. 7.1\%).\cite{bhimani2025_spanishcrc} After adjusting for demographic and call-level variables, Spanish-speaking patients remained twice as likely to opt in to screening (adjusted OR 2.012) (Bhimani et al., 2025). These findings challenge the assumption that AI disproportionately disadvantages non-English-speaking populations and instead suggest that language-concordant AI communication can meaningfully reduce screening disparities.
\subsection{Workflow Integration and Health-Plan Outcomes}Large-scale health-plan deployments demonstrate additional operational benefits. Organizations using the AI agent for outreach, care-gap closure, and member engagement reported higher connection rates and more consistent reach across eligible populations.\cite{homepage} These findings align with the Safety Framework’s emphasis on workflow integration, in which AI systems augment rather than replace staff by taking on repetitive communication tasks and providing stable capacity during periods of variable staffing. In the context of chronic disease monitoring, increased engagement observed in the nephrology remote-monitoring cohort further demonstrates how improved communication consistency can support downstream adherence behaviors.\cite{agnew2025_welcome}
\subsection{System-Wide Deployment and Operational Efficiency}The transition from pilot studies to enterprise-wide implementation represents a critical inflection point for healthcare AI. A 13-month prospective implementation study at WellSpan Health—an integrated health system serving south-central Pennsylvania and northern Maryland—provides the first comprehensive evidence of autonomous AI deployment as a system-wide capability rather than an isolated intervention.\cite{nejm}
From September 2024 through September 2025, the AI voice assistant ("Ana") conducted nearly 2 million patient conversations across three strategic deployment categories: targeted outbound campaigns for care gap closure, integrated automated outreach for routine patient communications, and inbound call management for patient-initiated inquiries.\cite{nejm} This scale of deployment—unprecedented for an autonomous healthcare AI agent—demonstrates the feasibility of AI integration as an enterprise capability.
\subsubsection{Procedural Preparation and Patient Education}In a colonoscopy preparation pilot, the AI agent contacted 1,627 patients who opted in to receive preparation coaching through a series of structured calls. Among these patients, 30.8\% (95\% CI: 28.6\%–33.0\%) completed full conversations with the agent. Patient experience metrics were notably strong: among 460 patients providing satisfaction ratings, 65.9\% (95\% CI: 61.5\%–70.3\%) rated likelihood-to-recommend as 9 or 10 on a 10-point scale, with a mean rating of 8.65.\cite{nejm} Qualitative analysis of patient feedback identified recurring themes of perceived empathy, patience, and appreciation for unlimited question opportunities—characteristics typically associated with high-quality human clinical communication.
\subsubsection{Diagnostic Results Communication}For mammogram results delivery, the AI agent reached 11,000 patients with normal results requiring notification. The system successfully connected with 5,019 patients (45.6\%, 95\% CI: 44.7\%–46.6\%) and completed full conversations with 2,734 (24.9\%, 95\% CI: 24.1\%–25.7\%). Patient satisfaction and likelihood-to-recommend ratings both exceeded 9 on a 10-point scale. The average duration of the conversation of 3.3 minutes resulted in more than 350 hours of direct patient engagement for education on annual mammograms and scheduling assistance for subsequent screenings \cite{nejm}.
\subsubsection{Workforce Augmentation in Primary Care}The most substantial operational impact emerged in primary care call center deployment. WellSpan's primary care call centers had historically struggled with staffing shortages, operating at only 50\% practice coverage with significant patient wait times. Following AI implementation, the system achieved dramatic capacity expansion \cite{nejm}:
\begin{itemize}
\item{Practice coverage expanded from 50\% to 100\% of primary care practices}
\item{The AI agent managed more than 50\% of all phone-scheduled appointments}
\item{Weekly talk time averaged 850 hours, equivalent to the workload of 28 full-time call center specialists}
\item{Patient refusal to engage with the AI agent remained below 3\%} 
\item{Efficiency gains effectively doubled staff productivity, enabling the call center to manage twice the workload without additional human staffing}
\end{itemize}
The AI agent was initially deployed to handle routine requests—operating hours, directions, parking information—freeing staff for complex calls requiring human judgment. Subsequently, its role expanded to include scheduling acute and routine primary care appointments, demonstrating successful scope extension based on validated performance.
\subsubsection{Implementation Success Factors}
Several factors distinguished this enterprise implementation from typical pilot studies. The health system treated AI as a system-wide capability requiring governance standards rather than an isolated technology deployment. Multidisciplinary teams including nurses, physicians, administrators, and quality improvement specialists collaborated with the AI development team to map patient journeys, identify intervention points, create conversation scripts, and develop safety protocols for clinical escalation.\cite{nejm} Frontline staff involvement throughout design and implementation proved critical—staff created and reviewed test conversations, provided feedback to refine delivery, and developed escalation pathways that route clinical or urgent issues appropriately. This user-centered approach, combined with clinical oversight, enabled the transition from isolated pilots to sustained operational deployment at scale.
\subsection{Patient Experience and Satisfaction Outcomes}Across diverse clinical applications, the AI care agent consistently achieved patient satisfaction metrics comparable to or exceeding benchmarks for human-delivered care. This pattern of high satisfaction emerged with patients' awareness that they were interacting with an AI system.

In the WellSpan implementation, satisfaction scores ranged from 8.65 to above 9.0 on 10-point scales across all deployment categories.\cite{nejm} The colonoscopy preparation pilot achieved a mean likelihood-to-recommend score of 8.65/10, with nearly two-thirds of respondents (65.9\%) providing promoter-level ratings of 9 or 10. Mammogram results delivery maintained satisfaction and likelihood-to-recommend averages above 9/10.

Qualitative feedback from patients revealed several themes explaining high satisfaction with AI-mediated communication \cite{nejm}:
\begin{itemize}
\item{Perceived empathy: Patients frequently commented on the AI agent's empathetic tone and apparent understanding of their concerns, suggesting successful implementation of conversational design principles emphasizing warmth and acknowledgment.}
\item{Patience and availability: Unlike time-constrained human interactions, patients appreciated the AI agent's unlimited availability for questions without perceived time pressure—a characteristic documented in prior qualitative research on patient preferences for AI communication.\cite{trivedi}.}
\item{Consistency: The standardized yet personalized delivery ensured all patients received complete, accurate information regardless of when they called or which agent instance they reached.}
\end{itemize}
These satisfaction findings challenge assumptions that patients inherently prefer human communication for healthcare interactions. When AI systems are designed specifically for empathetic, patient-centered conversation, rather than transactional information exchange, patient acceptance and satisfaction can match or exceed traditional delivery models. The high satisfaction levels observed across diverse use cases (e.g., procedural preparation, diagnostic results, appointment scheduling) suggest this pattern generalizes across healthcare communication contexts.

\subsection{Summary of Impact}The accumulating evidence supports the conclusion that AI-mediated communication can enhance care delivery by augmenting rather than replacing human clinicians, expanding care-team capacity, ensuring consistent high-quality interactions for all patients, and—when properly validated—achieving safety standards that match or exceed human clinical communication. The framework demonstrates that treating interaction intelligence as a first-class safety variable, combined with rigorous pre-deployment validation, enables deployment of autonomous AI agents that improve outcomes across the quadruple aim of healthcare: better patient experience, improved population health, reduced costs, and enhanced clinician well-being through workload redistribution.
    \section{Related Work}
\label{related_work}
\subsection{Clinical evaluation of LLMs}
Traditional healthcare LLM benchmark evaluations were mostly hinged on static, offline benchmarks which usually materialized as multi-choice Q/A fashion or report summarization. While such datasets offer convenience and reproducibility, they systematically miss the dynamic signals including contextual drift, longitudinal consistency, interpersonal variability, tool-use, and uncertainty negotiation that arise in authentic interactions. These benchmarks span a range of formats including classification, question answering, text and code generation, but share a common design pattern: each example is a self-contained input (a note, report, question, or schema) paired with a single “correct” output, and models are scored with pointwise metrics such as exact match, F1, or LLM-jury scores. For example, MedCalc-Bench\cite{khandekar2024medcalc}, CLEAR\cite{lopez2025clinical}, Medec\cite{abacha2025medec}, EHRSHOT\cite{wornow2023ehrshot}, and ADHD-Behavior\cite{pillai2024measuring} / ADHD-MedEffects\cite{bannett2025applying} evaluate classification or computational reasoning from notes and EHR codes, asking models to detect conditions, compute risk or severity, or flag documentation errors. Knowledge-focused QA sets such as HeadQA\cite{vilares2019head}, MedBullets\cite{chen2025benchmarking}, MedQA\cite{jin2021disease}, MedMCQA\cite{pal2022medmcqa}, PubMedQA\cite{jin2019pubmedqa}, MedicationQA\cite{abacha2019bridging} test exam-style or snippet-grounded questions with multiple-choice or binary answers. A large family of generation benchmarks—DischargeMe\cite{xu2024discharge}, MedAlign\cite{fleming2023medalign}, ACI-Bench\cite{yim2023aci}, MIMIC-RRS\cite{chen2023toward}, MIMIC-IV-BHC\cite{aali2025dataset}, and MedDialog\cite{zeng2020meddialog}—measure how well models summarize notes, radiology reports, or conversations and produce treatment plans, discharge instructions, or empathetic counseling responses, typically via rubric-based LLM juries. Other datasets target operational and safety-adjacent tasks such as identifying PHI or privacy risk (MedConfInfo\cite{rabbani2024evaluation}, PrivacyDetection\cite{tse2025large}), proxy senders (ProxySender\cite{tse2025large}), hallucinations (MedHallu\cite{pandit2025medhallu}), or research-oriented code generation from natural language (EHRSQL\cite{lee2022ehrsql}). While these benchmarks are valuable for coverage and comparability across models, they all instantiate the static offline paradigm: the model is evaluated on frozen, de-identified artifacts, with no real-time interaction, feedback, or evolving context. Tasks are typically single-shot and decontextualized (e.g., one question, one note, one discharge summary), and success is reduced to matching a reference label or receiving a high rubric score on a single response. As a result, these datasets provide snapshots of task competence rather than measuring how a system behaves over multi-turn, safety-critical episodes: they do not capture longitudinal patient trajectories, dynamic clinical decision-making, tool use, knowledge drift, user misunderstanding, or the role of redundancy and cross-checking between agents.

\subsection{Non-clinical evaluation of LLMs}
In non-clinical domains, researchers have developed benchmarks to evaluate conversational quality, empathy, and emotional alignment in AI companions. One line of work focuses on empathetic response generation.  Rashkin et al. \cite{rashkin2019towards} introduced the EmpatheticDialogues dataset of 25k open-domain conversations grounded in emotional situations. Building on such datasets, large-scale empathy benchmarks have emerged. For example, Huang et al. \cite{2023emotionally} present EmotionBench, with 400+ carefully curated scenarios designed to elicit eight key emotions. They conducted human studies with 1,200+ participants to establish reference responses, against which seven modern LLMs (GPT-4, etc.) were evaluated. Another benchmark, Heart-to-Heart Talk (H2H-Talk), targets supportive counseling dialogues. Wang et al. \cite{wang2025h2htalk} assemble 4,650 simulated conversations covering therapy-style scenarios and long-term persona development. Evaluation of 50 companion models on H2H-Talk revealed persistent challenges in maintaining long-horizon coherence and remembering evolving user needs.
The findings showed that while LLMs often respond with an appropriate emotional tone, they still fall short of human-level emotional alignment – e.g. misreading situational nuance and failing to mimic human patterns of emotional change.

\subsection{Speech Understanding and Voice Systems}
 Our system builds on advances in speech recognition and generation, especially for robust, context-aware performance in voice conversations. In automatic speech recognition (ASR), prior work has addressed adapting to conversation context and diverse audio conditions. For example, Amazon Alexa’s ASR can leverage context from prior dialogue turns to bias transcription. Hardesty \cite{hardesty2022engineering} reports that when Alexa knows the expected names (e.g. after asking “Do you mean Meg Jones or Meg Bauer?”), it favors the correct name (“Bauer”) over a more common phonetic match (“power”), reducing errors in follow-up replies by ~26\%. Such contextual biasing and on-the-fly customization of language models have become standard in production voice assistants. Another challenge is short utterance robustness – ASR systems often struggle with very brief inputs (e.g. “yes”, “no”, or backchannel words). A recent study by Zolnoori et al. \cite{zolnoori2024decoding} finds that a state-of-the-art commercial ASR had word error rate ~86\% on patient utterances under 5 words, compared to 37\% on longer (>11 word) utterances. This disparity underscores the difficulty of transcribing fragments with little context, a pertinent issue for real-time dialogues. To improve overall ASR resilience, massively multilingual and weakly supervised training has gained traction. OpenAI’s Whisper model  \cite{radford2023robust} exemplifies this: trained on 680k hours of web audio in ~100 languages, Whisper achieves near human-level transcription accuracy and strong noise robustness in a zero-shot setting. Similarly, Google’s Universal Speech Model (USM) \cite{zhang2023google} pretrains on 12 million hours across 300+ languages, enabling a single model to handle speech from over 100 languages with competitive results. These large-scale ASRs provide a foundation for multilingual voice agents. On the generation side, producing empathetic and natural speech has been a focus for industry voice assistants. Amazon introduced controllable speaking styles for Alexa, allowing developers to trigger a “disappointed/empathetic” tone or an “excited/happy” tone in responses \cite{peters2019alexa}. Our work leverages these advances by using robust ASR (with contextual biasing and multilingual support) and expressive neural voices, ensuring the AI agent can hear and speak with high fidelity and empathy across diverse scenarios.

 \subsection{Inference and System Architecture} Deploying real-time voice assistants powered by large LLMs poses significant systems challenges. Prior research has explored distributed and efficient inference techniques to meet latency requirements. One approach is to maximize throughput via optimized batching and caching. For example, vLLM (Kwon et al. \cite{kwon2023efficient}) introduces a PagedAttention mechanism to manage memory for the LLM’s key-value cache more efficiently. This allows dynamic batching of requests without wasted memory, and even sharing of past context across requests. The result is a serving system that achieves up to 24× higher throughput than conventional HuggingFace-based inference pipelines under the same latency budget \cite{kwon2023vllm}. Caching mechanisms are especially beneficial in multi-turn settings: by persisting the LLM’s state from previous turns in GPU memory, one can skip recomputation of the prompt prefix and only decode new user input. This prefix reuse yields huge speedups – effectively amortizing the cost of the initial prompt over many turns. (In our system, we observe a similar effect: after the first turn, the LLM operates almost entirely on the incremental input, which is a much smaller computational load.) To keep latency low, researchers have also pursued model compression and acceleration. Techniques like knowledge distillation and quantization are widely used to shrink model size or speed up inference. Distilled or compact models (e.g. DistilBERT, 8-bit quantized LLaMA) can run faster and even cache entire response candidates for reuse. Finally, ensuring reliability in real-time systems has led to architectures with verification and guardrail layers. These can be additional LLMs or rule-based checkers that intercept the primary model’s output to verify facts, filter unsafe content, or inject necessary corrections. Such layered designs – incorporating caching, specialization, compression, and verification – are crucial to achieving a scalable, real-time LLM-based voice agent that is both responsive and trustworthy in practice.
     \section{Conclusion}
 \label{sec:conclusions}
 
    By grounding the system design for Polaris in real patient interactions and governed telemetry, we reliably improved safety, empathy, equity, and workflow outcomes at clinical scale to build the safest generative AI system for healthcare. Production telemetry drives the architecture. Elevating interaction micro-skills to safety variables, investing in modality-specific models (e.g., contextual ASR), and leveraging hardware-aware serving unlocks non-linear gains for voice in healthcare. Orchestration designed for real workflows, validated under RWE-LLM, turns conversational quality into reliable clinical and operational outcomes. The production-first approach links signals to solutions to impact, providing a repeatable path for continued progress.

    \section*{Acknowledgments}
    We thank partner health systems, clinicians, and patients whose feedback shaped this work, and the engineering and research teams who built the system.

    \end{document}